\documentclass[12pt,preprint]{aastex}

\newcommand{\eg}{{e.g.,}}
\newcommand{\etal}{et\ al.}
\newcommand{\hzrg}{{HzRG}}
\newcommand{\hzrgs}{{HzRGs}}
\newcommand{\scoo}{{WN~J0305+3525}}
\newcommand{\degree}{\ifmmode {^{\circ}} \else {$^{\circ}$}\fi}
\newcommand{\ergps}{\ifmmode {\rm\,erg\,s^{-1}} \else {${\rm\,erg\,s^{-1}}$}\fi}
\newcommand{\ergpspcm}{\ifmmode {\rm\,erg\,s^{-1}\,cm^{-2}} \else {${\rm\,erg\,s^{-1}\,cm^{-2}}$}\fi}
\newcommand{\Kz}{{$K-z$\ }}
\newcommand{\lya}{{\rm\,Ly$\alpha$}}

\newcommand{\LFIR}{\ifmmode {\,L_{\rm FIR}}\else ${\,L_{\rm FIR}}$\fi}
\newcommand{\Mdust}{\ifmmode {\,M_{\rm dust}}\else ${\,M_{\rm dust}}$\fi}
\newcommand{\Lsun}{\ifmmode {\rm\,L_\odot}\else ${\rm\,L_\odot}$\fi}
\newcommand{\Msun}{\ifmmode {\rm\,M_\odot}\else ${\rm\,M_\odot}$\fi}
\newcommand{\Rsun}{\ifmmode {\rm\,R_\odot}\else ${\rm\,R_\odot}$\fi}
\newcommand{\OmM}{\ifmmode {\Omega_{\rm M}}\else $\Omega_{\rm M}$\fi}
\newcommand{\OmL}{\ifmmode {\Omega_{\Lambda}}\else $\Omega_{\Lambda}$\fi}
\newcommand{\kmps}{\ifmmode {\rm\,km~s^{-1}} \else ${\rm\,km\,s^{-1}}$\fi}
\slugcomment{Preliminary version for ApJ}

\shorttitle{Obscured Radio Galaxies at High Redshift}
\shortauthors{Reuland \etal}

\received{2002 September 13}
\begin{document}

\title{An Obscured Radio Galaxy at High Redshift}

\author{Michiel Reuland,\altaffilmark{1,2,3} Wil van
Breugel,\altaffilmark{1} Huub R\"ottgering,\altaffilmark{2} Wim de
Vries,\altaffilmark{1} Carlos De Breuck\altaffilmark{4} \& Daniel
Stern\altaffilmark{5}}

\altaffiltext{1}{Institute of Geophysics and Planetary Physics, 
L$-$413 Lawrence Livermore National Laboratory, P.O. Box 808, 
Livermore, CA 94551, U.S.A., email: mreuland@igpp.ucllnl.org}  
\altaffiltext{2}{Leiden Observatory, P.O. Box 9513, 2300 RA
Leiden, The Netherlands} 
\altaffiltext{3}{Department of Physics, UC Davis, 1 Shields Avenue, 
Davis, CA 95616, U.S.A.}
\altaffiltext{4}{Institut d'Astrophysique de Paris, CNRS, 98bis
Boulevard Arago, F-75014 Paris, France}
\altaffiltext{5}{Jet Propulsion Laboratory, California Institute of
Technology, Mail Stop 169-327, Pasadena, CA 91109, U.S.A.}

\begin{abstract}
\noindent Perhaps as many as 10\% of high redshift radio galaxy
(\hzrg; $z > 2$) candidates that are selected using an Ultra Steep
radio Spectrum (USS) criterion fail to show optical emission
(continuum, lines) in deep Keck exposures. Their parent objects are
only detected in the near-IR and are probably heavily obscured and/or
at very high redshift.  To search for signatures of dust and help
constrain the nature and redshifts of these ``no-$z$'' radio galaxies,
we have conducted a program of submillimeter and millimeter
observations.  Here we report the first results of a detailed study of
one of these objects, \scoo.

\scoo\ appears associated with a small group of $K \sim 21 - 22$
objects and is strongly detected at both 850\,\micron\ and 1.25\,mm.
On the basis of its faint $K$-band magnitude, spectral energy
distribution (SED) and other evidence we estimate that the radio
galaxy is probably at a redshift $z \simeq 3 \pm 1$. This would make
\scoo\ a radio-loud Hyper Luminous Infrared Galaxy ($L_{\rm FIR} \sim
10^{13} \Lsun$) similar to, but more obscured than, other dusty radio
galaxies in this redshift range.  This, together with the absence of
\lya\ emission and compact ($\theta \lesssim 1\farcs9$) radio
structure, suggests that \scoo\ is embedded in a very dense, dusty
medium and is probably at an early stage of its formation.

\end{abstract}


\section{Introduction}

High redshift active galactic nuclei (AGNs) have gained renewed
interest as cosmological probes with the findings that (i) there is a
good correlation between the masses of the stellar bulges
of galaxies and their central black holes suggesting a causal
connection in their formation \citep[\eg][]{Magorrian98aj} and (ii)
that a population of high redshift heavily obscured AGNs seems the
best candidate to account for a substantial fraction (30$-$40\%) of
the 5$-$10 keV cosmic X-ray background (XRB)
\citep[\eg][]{Hasinger02astroph,Stern02apj,Willott02mnras}.

The most vigorously star forming galaxies radiate strongly in the
far-IR \citep[][]{SandersMirabel96araa}. Sub-mm observations
show that as many as 50\% of $z > 3$ \hzrgs\ are ``Hyper Luminous
Infrared Galaxies'' \citep[HyLIRGs, $L_{\rm FIR} \sim 10^{13} \Lsun$;
Reuland \etal\ in prep.;][]{Archibald01mnras}.  While some HyLIRGs are
powered by an obscured AGN, there also exist many in which the
starburst dominates \citep{RowanRobinson00mnras}.  The detection of
strong rest-frame UV absorption features show that there are hot young
stars at $z \sim 3.8$ in at least one case \citep{Dey97apj}, and CO
mm-interferometry studies show that star formation occurs galaxy wide
over scales of up to $\sim$ 30\,kpc \citep{Papadopoulos00apj}.  These
observations provide direct evidence that powerful \hzrgs\ are massive
galaxies forming stars at rates as high as 2000\,$\Msun\ \rm yr^{-1}$.
From the near-IR Hubble \Kz diagram \hzrgs\ are known to be the most
luminous galaxies at any redshift up to $z \sim 5.2$
\citep[][hereafter DB02]{DeBreuck02aj}. Furthermore,
actively-accreting supermassive black holes \citep[$M_{\rm BH} \sim
10^{9} \Msun$;][]{Lacy01apj} are required to power radio sources.
Therefore, \hzrgs\ are excellent targets to study the formation and
coevolution of the most massive galaxies and their black holes.

\citet{DeBreuck01aj} found that 36\% of a sample of 33 USS
($\alpha_{325 \rm MHz}^{1400 \rm MHz} \lesssim -1.30$; $S_{\nu}
\propto \nu^{\alpha}$) selected radio sources with spectroscopic
information was at redshifts $z > 3$.  A surprisingly large fraction
(24\%) of these USS sources did {\it not} show emission lines even in
moderately long exposures (1$-$2\,hrs) at Keck. Of these latter
sources, 3 (or 9\% of the entire sample) did not show optical
continuum emission down to a limit of $R \sim 25$ and were detected only
at near-IR ($K$-band) wavelengths.

The nature of these ``no-$z$'' objects is unclear.  As discussed by De
Breuck \etal\ one can think of the following possible explanations:
objects which do show optical continuum (i) may be in the ``redshift
desert'' ($1.5 < z < 2.3$) where strong emission lines fall outside
the observable window or (ii) may be pulsars; objects without optical
continuum (iii) may be highly obscured AGNs or (iv) be at such high
redshift ($z \gtrsim 7$) that \lya\ is shifted out of the optical
passband.  In the latter two cases they may be young \hzrgs\ in an
exceptionally vigorous stage of their formation, and one would expect
large amounts of dust associated with them. To search for this we have
conducted a program of sub-mm and mm observations of these ``no-$z$''
radio galaxies. Here we report the first results of a detailed study
of one such object, \scoo.

To facilitate comparison with other papers we adopt an universe
with $\OmM = 1.0$, $\OmL = 0$, and $H_{0} = 50 \kmps\,\rm Mpc^{-1}$
unless stated otherwise.  The angular scale at $z = 3$ is then
7.3\,$\rm kpc~arcsec^{-1}$ and the look-back time is 11.4\,Gyr, or
88\% of the age of the universe.

\section{Observations and Results}

\subsection{Selection and Keck Imaging} 

VLA observations by \citet{DeBreuck00aas} show that \scoo\ is a
compact ($\theta \lesssim 1\farcs9$) steep spectrum ($\alpha_{325 \rm
MHz}^{1.4 \rm GHz} = -1.33$) radio source with flux density $S_{\rm
1.4 GHz} = 15.8$\,mJy.  As part of our program aimed at finding $z >
3$ radio galaxies among USS sources DB02 obtained a 15\,min $I$-band
image with 0\farcs7 seeing and a 38\,min $K$-band image with 0\farcs55
seeing at Keck.  We obtained a 58\,min Keck $J$-band image on UT 2000
January 30 (seeing $\sim 0\farcs6$) with the Near Infrared Camera
\citep[NIRC;][]{MatthewsSoifer94proc}, which was reduced following
DB02.  Formal 5$\sigma$ detection limits in 2\arcsec\ diameter
circular apertures are $I_{\rm lim} = 23.37$, $J_{\rm lim} = 23.15$,
and $K_{\rm lim} = 22.61$.  Astrometry for the $I$-band image was
determined using the USNO-A2.0 catalog \citep{Monet98aas} giving a
formal uncertainty of 0\farcs15.  The absolute rms uncertainties with
respect to the international celestial reference frame of the catalog
and radio image are 0\farcs2 and 0\farcs25 respectively (\eg\ DB02).
The small near-IR images were registered to the $6 \arcmin \times
7\arcmin$ $I$-band image to better than 0\farcs05 rms, resulting in
relative astrometry between the near-IR and radio images accurate to
$1 \sigma \sim 0\farcs4$ rms.

\begin{figure}[t]
\plotone{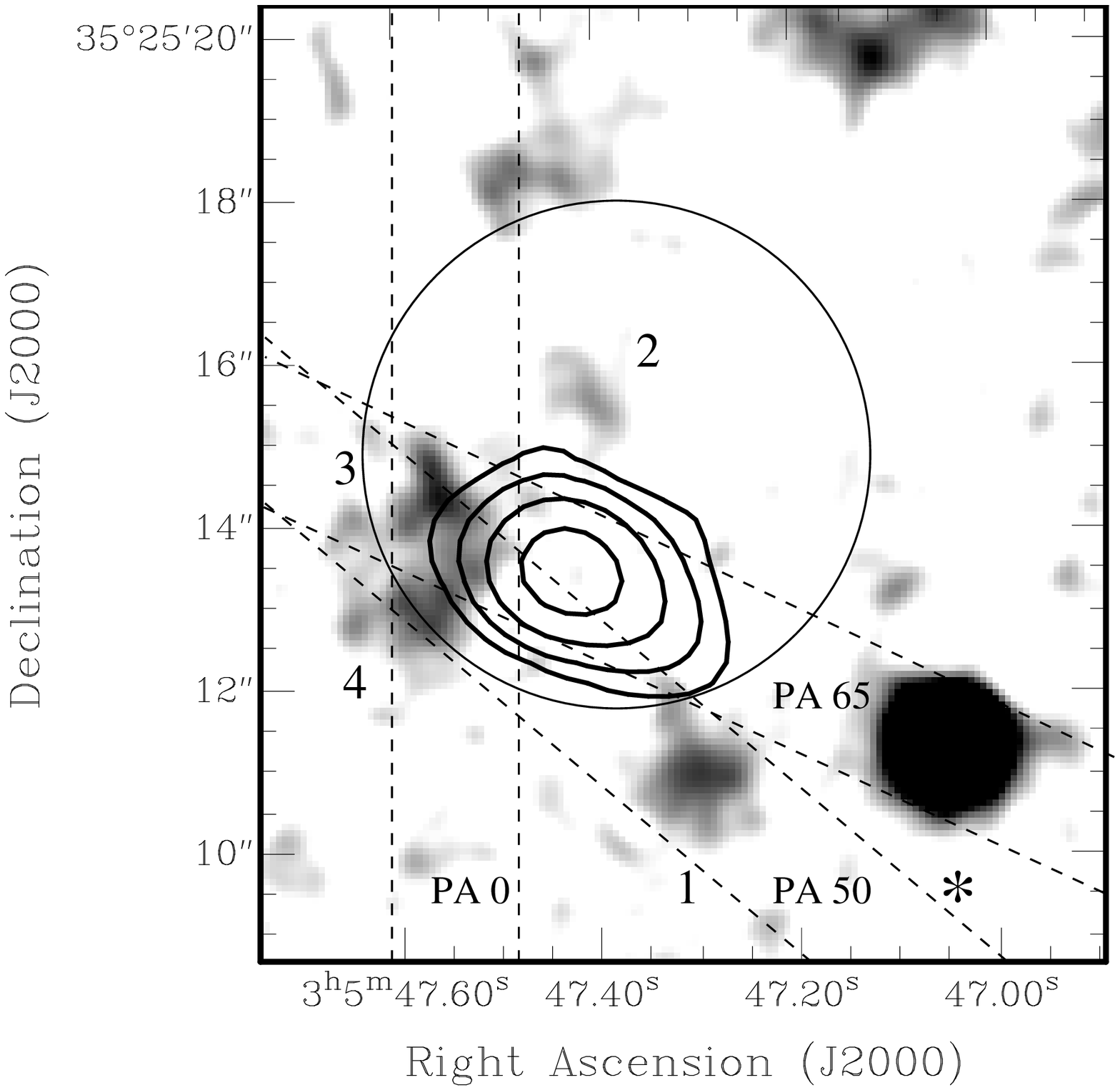}
\caption{\small Keck/NIRC $K$-band image with 4.85\,GHz VLA radio
contours \citep{DeBreuck00aas} overlaid of the \hzrg\ candidate
\scoo. The greyscale image has been smoothed to a resolution of
0\farcs7 and the contour levels are 0.25, 0.5, 1, and
2\,mJy\,beam$^{-1}$.  Note the multiple components. The bright object
to the SW is a spectroscopically confirmed star. The circle with
3\arcsec\ radius represents the nominal 3$\sigma$ astrometric
uncertainty for the centroid of the 850\,\micron\ emission and dashed
lines indicate the LRIS slit positions.
\label{0305K}}
\end{figure}

\begin{figure}[t]
\plotone{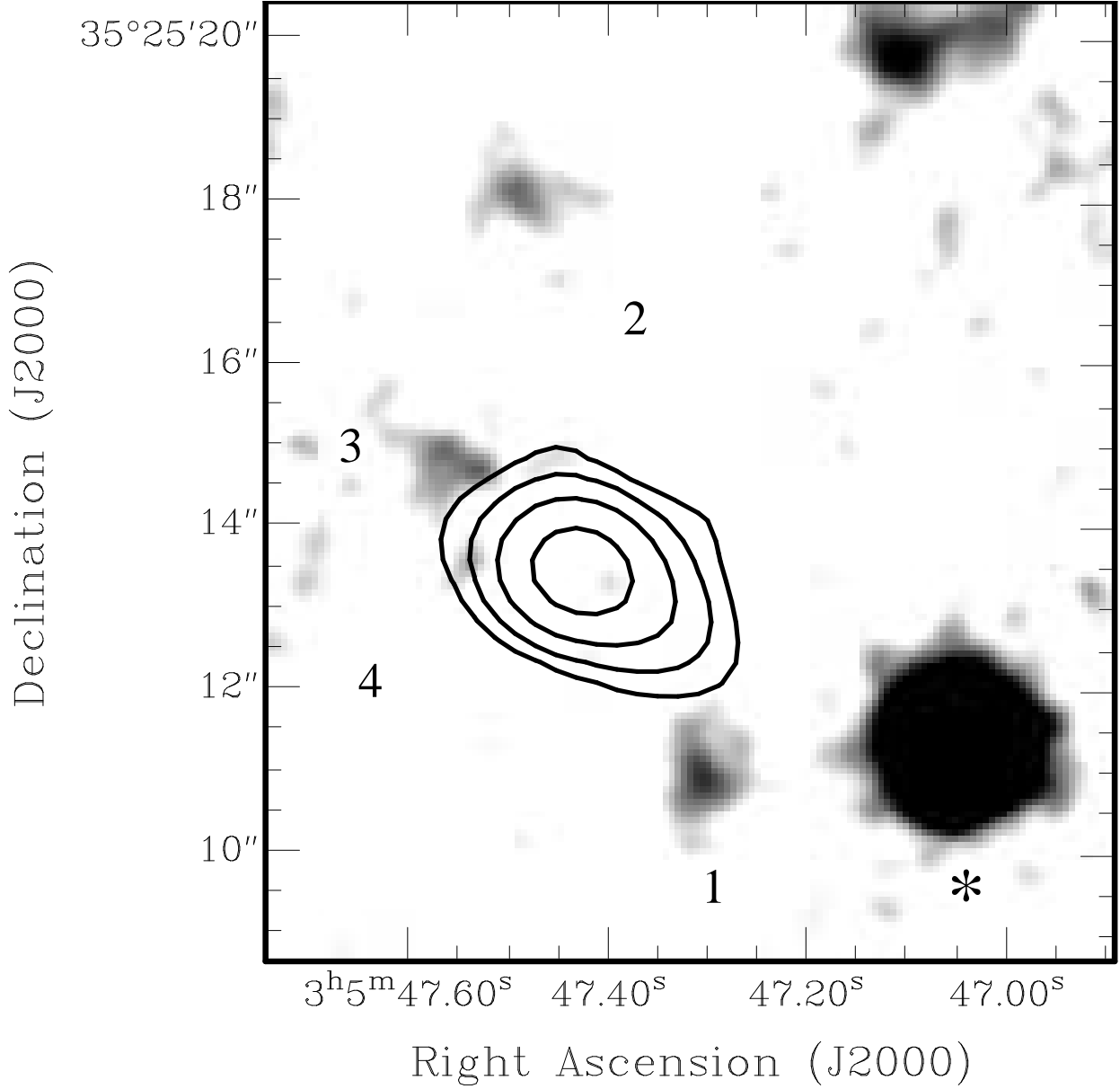}
\caption{\small Same as Figure \ref{0305K}, but for the greyscale
representation of the Keck/NIRC $J$-band image of \scoo.  The image
has been smoothed to a resolution of 0\farcs7. Objects 2 and 4 which
were visible in $K$-band remain undetected. This image is similar
to the $I$-band image (not shown).
\label{0305J}}
\end{figure}

Figure \ref{0305K} shows the $K$-band image with 5\,GHz radio map
overlaid.  It reveals fuzzy structure with multiple components within
2\arcsec$-$3\arcsec\ of the radio source. Note that none of the
objects appears directly associated with the radio emission itself.
Table \ref{table} summarizes the observed properties of these
objects. Objects 2 and 4 remain undetected in the $I$ and $J$-band
images (Fig. \ref{0305J}).

\subsection{JCMT and IRAM Observations} We observed \scoo\ on UT
1999 December 6 and 7 at 850\,\micron\ and 450\,\micron\ with the
Submillimetre Common-User Bolometer Array
\citep[SCUBA;][]{Holland98spie} at the James Clerk Maxwell Telescope
in stable atmospheric conditions $\tau_{850} \sim 0.29 - 0.32$. We
performed two sets of 50 integrations each night using the 9-point
jiggle photometry mode while chopping 45\arcsec\ in azimuth at
7.8\,Hz.  Pointing was better than 2\arcsec. Flux calibration was
performed on Saturn, HLTAU, and CRL618 and we adopted a gain $C_{850}
\sim$ 182\,mJy\,beam$^{-1}$\,V$^{-1}$. The data were reduced following
procedures outlined in the SCUBA Photometry Cookbook\footnote{The
SCUBA Photometry Cookbook is available at
http://www.starlink.rl.ac.uk/star/docs/sc10.htx/sc10.html}.  Combining
the data from both nights yields an average 850\,\micron\ flux density
of 12.5 $\pm$ 1.5\,mJy, while at 450\,\micron\ we measured 21 $\pm$
19\,mJy and did not detect the object. The FWHM of the beam at
850\,\micron\ for SCUBA is $\theta_{\rm FWHM} = 14\farcs5$.  We imaged
\scoo\ with SCUBA on UT 2000 February 26, 27, and 28, reaching $\sim
1.5$\,mJy rms, to obtain a more accurate position for the sub-mm
source. The position of the 850\,\micron\ centroid $\alpha = 03^{\rm
h}05^{\rm m}47\fs 38$, $\delta = +35 \degree 25\arcmin 15\farcs0$
(J2000) with estimated 3$\sigma$ astrometric uncertainty of $\pm\, 3
\times \theta_{\rm FWHM}/(2 \times S/N) \sim 3 \arcsec$
\citep[following][]{Serjeant02astroph} is indicated in Figure
\ref{0305K}.

\scoo\ was also observed at 1.25\,mm using the 37 channel MPIfR
Bolometer Array \citep[MAMBO II;][]{Kreysa98spie} at the IRAM 30\,m
telescope on UT 2000 July 18. We performed 11 symmetric ``ON-OFFs''
with a chop-nod distance of 45\arcsec, each with 20 subscans of 10\,s
per subscan. The atmospheric extinction $\tau_{1.25}$ varied between
0.16 and 0.34 and pointing corrections were less than 2\arcsec. The
calibration factor of 13.3 counts\,mJy$^{-1}$ was estimated from
on-offs on Uranus and Saturn.  Using MOPSI with sky-noise filtering we
detect the object at 4.17 $\pm$ 0.56 mJy.

\subsection{Keck Spectroscopy} 
We have attempted to measure the redshift of \scoo\ using the
Low-Resolution Imaging Spectrometer \citep[LRIS;][]{Oke95pasp} with
the $150\, \ell \,\rm mm^{-1}$ grating blazed at 7500\,\AA\ and
1\farcs5 wide slit resulting in a spectral resolution of 15\,\AA\ and
wavelength coverage of $4100 - 10000$\,\AA.  Between exposures we
shifted the object by 10\,\arcsec\ along the slit to facilitate fringe
removal in the red parts of the CCD.  The slit positions are indicated
in Figure \ref{0305K}.  A 3 $\times$ 30\,min observation at
PA=50\degree\ on UT 1999 December 20 through targets 1 and 3 did not
show emission lines.  On UT 2000 February 1 we first observed 2
$\times$ 30\, min at PA=65\degree\ through target 3, the formal radio
source position, and the SW stellar object ($*$). No emission lines
were found, but the latter object was identified as a Galactic M
giant. A second set of similar exposures at PA=0\degree\ through
targets 3 and 4 did not show emission lines either, but we may have
detected faint continuum ($F_{\lambda}^{\rm cont} \approx 2 \pm 3
\times 10^{-19} \ergpspcm$\,\AA$^{-1}$) redward of $\lambda \approx
6300$\,\AA\ from object 3.  Object 2 was not observed
spectroscopically, but it is unlikely that it would have been detected
given its faintness.  Assuming that \lya\ would fall between
5100\,\AA\ and 8900\,\AA\ where the sky is well behaved and assuming a
line width $\sim$30\,\AA\ \citep[\eg][]{Dey97apj}, we derive a
$3\sigma$ upper limit to the \lya\ flux density of $3 \times 10^{-17}
\ergpspcm$.

\section{Discussion} 

\subsection{Identification and Redshift Estimate}

What are the chances that the sub-mm source, the radio source and the
$K$-band objects are all unrelated? Sub-mm surveys have shown that the
$S_{850\,\micron} \gtrsim 10$\,mJy number density is less than 100 per
square degree \citep{Hughes02mnras}. Radio surveys show that there are
less than 12 radio sources with $S_{\rm 1.4 GHz} \gtrsim 15$\,mJy per
square degree \citep{deVries02aj}. Within the sub-mm error circle
(3\arcsec\ radius; by far the largest of the three), the chance of
finding a chance superposition of a $\gtrsim 10$\,mJy sub-mm source
and a $\gtrsim 15$\,mJy radio source is negligible ($< 2.3 \times
10^{-4}$).  The chance of finding three unrelated $K < 22$\,mag
objects within the error circle is less than 10\% given a number
density $\sim 2 \times 10^{5}$ per square degree
\citep{Djorgovski95apj}. Despite the offsets between the $K$-band
objects and the radio source there are good reasons to believe that
they are related. Many USS selected radio galaxies consist of multiple
components \citep{vanBreugel98apj}. Furthermore, in most \hzrgs\ the
radio AGN are obscured \citep{Reuland02prepb} and there are other
cases where the AGN appears to be located off center from the bulk of
the parent galaxy \citep[\eg\ 3C294;][]{Quirrenbach01apj}.

Without a spectroscopic redshift we can use only indirect arguments to
estimate the redshift of \scoo\ such as the \Kz diagram, the shape of
the SED and the near-IR colors.  We cannot use the radio to sub-mm
index as a redshift indicator \citep{CarilliYun99apj} as \scoo\ is
radio loud.  The \Kz diagram as determined for \hzrgs\ (DB02) seems to
hold also for the sub-mm population \citep{Dunlop02astroph}.  If we
take a total magnitude of $K \sim 21$ to be representative for \scoo\
(Table \ref{table}), then we estimate from the \Kz diagram that \scoo\
is at $3 \lesssim z \lesssim 7$.

\begin{figure}[t]
\plotone{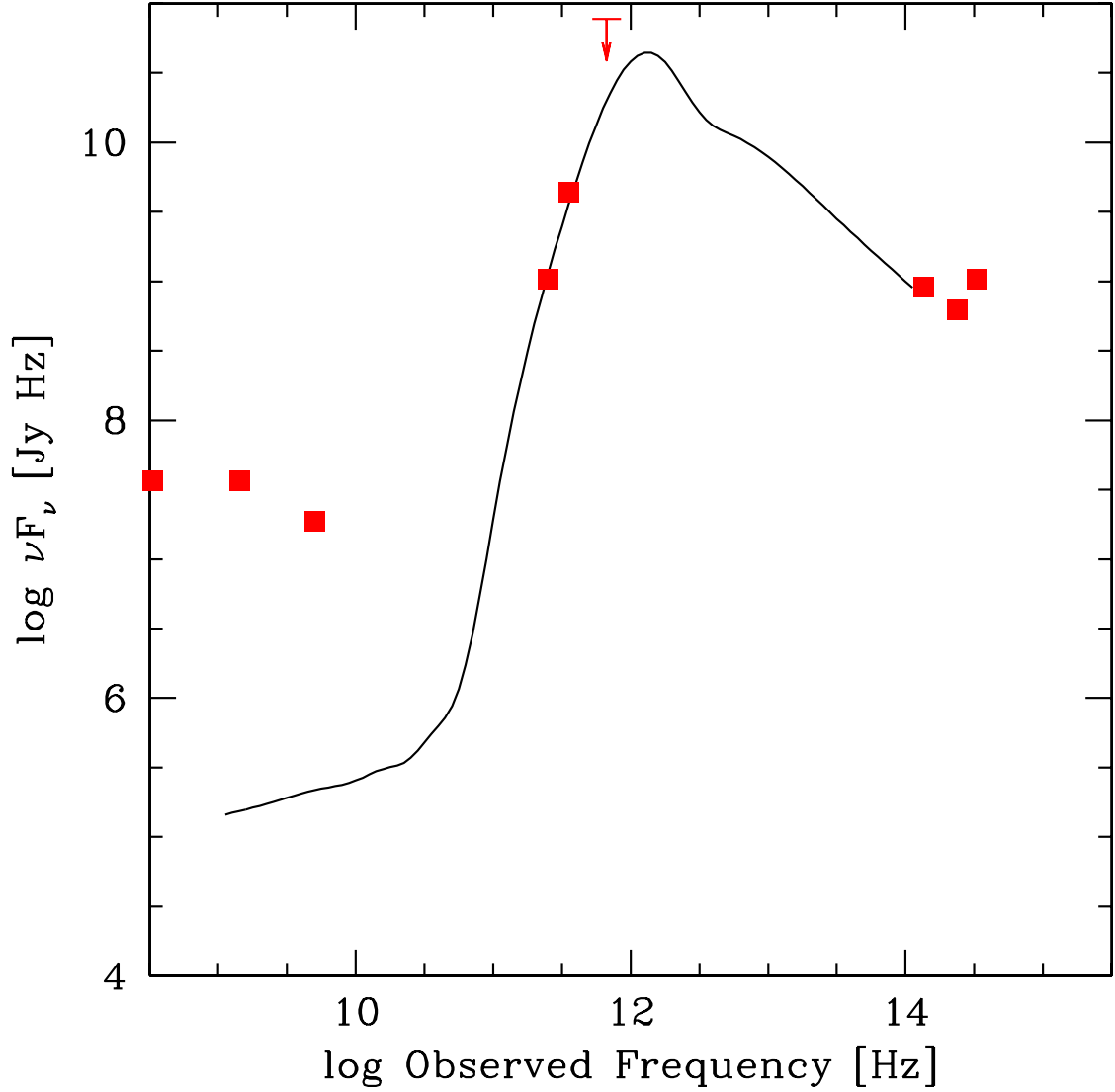}
\caption{\small Radio to near-IR SED of \scoo\ using data from the VLA
\citep{DeBreuck00aas}, IRAM, JCMT, and Keck (squares and $5\sigma$
upper limit for 450 micron point). Extrapolation of the steep radio
spectrum shows that the contribution of the non-thermal radio emission
is negligible at sub-mm wavelengths.  For comparison we overplot a
schematic SED representing local ultraluminous galaxies, shifted to
$z=3$ \citep[][$H_{0} = 75 \kmps\,\rm Mpc^{-1}$]{Lutz01aa}.
\label{0305SED}}
\end{figure}

Comparing the (sub)-mm ratios $S_{1250}/S_{850} = 0.33$ and
$S_{450}/S_{850} \lesssim 6$ with redshifted models of dusty
starforming galaxies \citep[c.f. Fig. 3 in][]{Hughes98nat} results in
redshift estimates of $3 \lesssim z \lesssim 6$ and $z > 1 - 2$
respectively. Moreover, almost all USS radio galaxies that are
detected in the sub-mm are at redshifts $z > 2.5$.  Figure
\ref{0305SED} represents the observed SED of \scoo\ and shows that the
(sub-)mm emission is of thermal origin and effectively rules out that
the object is a pulsar. A starburst SED with $1 \lesssim z \lesssim 4$
would fit the data closely at both the near-IR and (sub-)mm points.

The colors of the $K$-band objects are too uncertain to usefully
constrain the redshifts using stellar evolution models
\citep{StevensLacy01mnras}. The possible detection of faint continuum
redward of $\lambda \approx 6300$\,\AA\ in object 3 is consistent with
the object being at high redshift: it could be featureless continuum
redward of a \lya\ forest region at $z \sim 4.2$. 

Based on the above arguments we conclude that there is very good
circumstantial evidence that \scoo\ is at high redshift, probably at $z
\simeq 3 \pm 1$.

\subsection{Implications}

If \scoo\ is indeed at $z \simeq 3$ then its restframe FIR properties
are \LFIR\ $\sim 10^{13} \Lsun$, implying \Mdust\ $\sim 1.5 \times
10^8 \Msun$ and star formation rate $\sim 1500 \Msun\ \rm yr^{-1}$ for
standard dust models and cosmology \cite[c.f.  4C~60.07; $H_{0} = 75
\kmps\,\rm Mpc^{-1}$;][]{Papadopoulos00apj}.  This is comparable to
the radio-``quiet'' sub-mm source Lockman~850.1 which has been
estimated to be at $2 \lesssim z \lesssim 4$, based on its sub-mm/mm
SED \citep{Lutz01aa}, as well as other, radio-``loud'' galaxies in
this redshift range.  Compared with traditional \hzrgs, the \lya\
emission ($L_{\rm Ly\alpha} \lesssim 10^{42.3} \ergps$) is relatively
weak given its radio power ($P_{325MHz} \simeq 10^{35.1} \ergps\,\rm
Hz^{-1}$), but still within the large observed scatter
\citep[c.f. Fig. 10 in][]{DeBreuck00aa}. Thus \scoo\ appears to be a
heavily obscured radio-loud luminous sub-mm source.  We consider two
options that could explain the strong obscuration.

Recently, \citet{Greenberg00astroph} and \citet{Todini01mnras}
suggested that dust particles in high redshift galaxies may be
predominantly smaller than in local starbursting systems. The argument
is based on a different origin of the dust: at high redshifts there
has been little time for M giants (the primary sources of local dust)
to appear, and most early dust would be produced by M supergiants and
supernovae instead. The increase of the sub-mm detection rate with
redshift for radio galaxies
\citep[Reuland \etal\ in prep.;][]{Archibald01mnras} 
might then be interpreted as being due to changing dust properties
with redshift.  Smaller dust particles lower the estimated dust masses
and star formation rates and, because they are more efficient UV
absorbers, this might explain why some of the most massive starforming
galaxies are not seen in the optical at all.

Another reason that \scoo\ is heavily dust enshrouded might be related
to its young evolutionary stage. This is supported by the compactness
of the associated radio source for \scoo\ and the other ``no-$z$''
\hzrg-candidates. It seems reasonable to assume that, on average,
compact sources are younger than more extended ones
\citep{Blundell99aj}.  The dense circumnuclear dust configuration
would then quench the \lya\ emission.  

This suggests a scenario in which massive radio galaxies form with a
large starburst,
perhaps simultaneously in several smaller, merging components.
This starburst 
manifests itself as a luminous ``sub-mm source''.  During the
next stage a massive black hole is formed and/or activated, but any
rest-frame optical emission remains obscured by dust from the
starburst.  As the radio source evolves it grows and expels the
obscuring dust envelope, becoming visible as a \hzrg\
\citep[c.f.][]{Jarvis02prep,Reuland02prepb}.  This is essentially the
radio-loud version of scenarios where ``Ultra Luminous Infrared
Galaxies'' evolve into quasars as envisaged by \citet{Sanders88apj}.

Radio galaxies are the most massive and most extended galaxy sized
systems that are known to exist at high redshift. They are therefore
good targets for studying the origin of the relationship between
galaxies and their central black holes. \scoo\ may be at the stage
where both the galaxy and the black hole are being put together. As
with other sub-mm sources it is difficult to obtain spectroscopic
redshifts and it is important to obtain further information about
their SEDs at mid- and far-IR wavelengths.  X-ray imaging could show
whether or not there is a bright AGN associated with \scoo.  We are
actively pursuing follow-up observations to address some of these
issues.

\acknowledgments

We gratefully acknowledge the help of the Keck, and JCMT staff. We
thank the IRAM staff and Albrecht Sievers in particular for carrying
out observations in Director's Discretionary Time, project Delta
00-04, and reducing the data.  The work of M.R., W.v.B., and
W.d.V. was performed under the auspices of the U.S. Department of
Energy, National Nuclear Security Administration by the University of
California, Lawrence Livermore National Laboratory under contract
No. W-7405-Eng-48. The work of D.S. was carried out at the Jet
Propulsion Laboratory, California Institute of Technology, under a
contract with NASA.

\clearpage

\begin{deluxetable}{lrrrrrcc}
\tabletypesize{\scriptsize}
\tablecaption{Optical, near-IR, 850\micron\ and radio parameters of \scoo. \label{tbl-1}}
\tablehead{
\colhead{Source} & \colhead{R.A. (J2000)} & \colhead{DEC (J2000)} & \colhead{$I$\tablenotemark{a}} & \colhead{$J$\tablenotemark{a}} & \colhead{$K$\tablenotemark{a}} & \colhead{$(I-J)$\tablenotemark{b}} & \colhead{$(J-K)$\tablenotemark{b}} 
}
\startdata
1	 & 3:05:47.30 & +35:25:11.0 & 22.88  $\pm$ 0.15	& 22.03  $\pm$ 0.20	& 21.16  $\pm$ 0.19 & 0.9 & 0.9 \\
2	 & 3:05:47.40 & +35:25:15.6 & $> 23.37$  	& $> 23.15$	& 22.18  $\pm$ 0.33 & \nodata & $\gtrsim$ 1.0 \\
3	 & 3:05:47.56 & +35:25:14.4 & 23.16  $\pm$ 0.19	& 22.06  $\pm$ 0.20	& 21.30  $\pm$ 0.20 & 1.1 & 0.8\\
4	 & 3:05:47.56 & +35:25:13.0 & $> 23.37$     	& $> 23.15$	& 21.44  $\pm$ 0.24 & \nodata & $\gtrsim$ 1.7 \\
Ensemble\tablenotemark{c} & & & 23.41 $\pm$ 0.92	& 21.07  $\pm$ 0.18	& 20.71  $\pm$ 0.24 & $\gtrsim 2.3$ & 0.4 \\
850\,\micron & 3:05:47.38 & +35:25:15.0 \\  
4.85\,GHz\tablenotemark{d} & 3:05:47.42 & +35:25:13.4 \\  
SW Stellar Object ($*$)	 & 3:05:47.05 & +35:25:11.3  & 19.61  $\pm$ 0.01	& 18.50  $\pm$ 0.04	& 18.36  $\pm$ 0.05 & 1.1 & 0.1 \\
\enddata
\tablenotetext{a}{Magnitudes have been corrected for Galactic
reddening using the \citet{Cardelli89apj} extinction curve with $A_{V}
= 0.85$ as determined from the IRAS 100\,$\mu$m maps of
\citet{Schlegel98apj} and were measured in 2\arcsec\ diameter circular
apertures, unless noted otherwise.}
\tablenotetext{b}{Calculated using formal 5$\sigma$ detection limits
in the aperture ($I_{\rm lim} = 23.37$, $J_{\rm lim} = 23.15$, $K_{\rm
lim} = 22.61$) if fainter than those.\label{table}}
\tablenotetext{c}{Magnitudes have been measured in a 6\arcsec\ diameter
circular aperture centered on object 3 also encompassing objects 2
and 4.}
\tablenotetext{d}{Position from \citet{DeBreuck00aas}}
\end{deluxetable}

\clearpage

\clearpage

\clearpage

\end{document}